\documentclass[prd, twocolumn,floats,floatfix,nofootinbib,10pt] {revtex4-1}
\usepackage{graphicx}
\usepackage{dcolumn}
\usepackage{bm}
\usepackage{graphics}
\usepackage{slashed}
\usepackage{amssymb}
\usepackage{natbib}
\usepackage{amsmath}
\usepackage{url}
\usepackage[usenames,dvipsnames,svgnames,table]{xcolor}
\newcommand{\bea}{\begin{eqnarray}}
\newcommand{\ena}{\end{eqnarray}}
\newcommand{\bean}{\begin{eqnarray*}}
\newcommand{\enan}{\end{eqnarray*}}

\begin{document}

\title{On the geometrical hypotheses underlying wave functions\\ and their emerging dynamics}% Force line breaks with \\

\author{Erico Goulart}
\homepage{egoulart@ufsj.edu.br}
\affiliation{Physics Department, Universidade Federal de S\~ao Jo\~ao d'El Rei, C.A.P. Rod.: MG 443, KM 7, CEP-36420-000, Ouro Branco, MG, Brazil}

\author{Nelson Pinto-Neto}
\homepage{nelsonpn@cbpf.br}
\affiliation{Centro Brasileiro de Pesquisas F\'{\i}sicas, Rua Dr. Xavier Sigaud 150, Urca, CEP-22290-180, Rio de Janeiro, RJ, Brazil}

\date{\today}% It is always \today, today,
             %  but any date may be explicitly specified

\begin{abstract}
Classical mechanics for individual physical systems and quantum mechanics of non-relativistic particles are shown to be exceptional cases of a generalized dynamics described in terms of maps between two manifolds, the source being configuration space. The target space is argued to be 2-dimensional and flat, and their orthogonal directions are physically interpreted. All terms in the map equation have a geometrical meaning in the target space, and the pull-back of its rotational Killing one-form allows the construction of a velocity field in configuration space. Identification of this velocity field with tangent vectors in the source space leads to the dynamical law of motion. For a specific choice of an arbitrary scalar function present in the map equation, and using Cartesian coordinates in the target space, the map equation becomes linear and can be reduced to the Schr\"odinger equation. We link the bi-dimensionality of the target space with the essential non-locality of quantum mechanics. Many extensions of the framework here presented are immediate, with deep consequences yet to be explored.
\end{abstract}

%Use showkeys class option if keyword
                              %display desired

\maketitle

\section{\label{sec:Int}Introduction}

Quantum theory offers a precise set of rules which allows scientists to calculate, and predict with absolute experimental success, an immense variety of probabilistic results, in all energy and length scales already accessible for human scrutiny, from non-relativistic particles to relativistic quantum field theory, from the hydrogen atom to the Higgs field. However, the pillars on which these set of rules are constructed are still under intense debate. The standard view, the Copenhaguen interpretation \cite{Cop1,Cop2}, is not anymore a consensus, and has been severely criticized, essentially due to its imprecision (e.g. the fuzzy border between the quantum and the classical, the measurement problem, necessary reference to observers) \cite{Schr,Bell}, and its difficulty to furnish a good point of departure for future developments (like being extended to be applied to the whole Universe) \cite{Everett,Nelson}. Because of this, many alternatives have been proposed: the many worlds interpretation \cite{Everett,DeWitt}, the de Broglie-Bohm quantum dynamics \cite{Bohm,Holland,Nelson}, spontaneous collapse theories \cite{Pearle,Ghirardi1,Ghirardi2}, the consistent histories approach \cite{Griffths,Omnes,Hartle}, and so on. However, none of the proposals acquired consensus in the Physics community.

In order to shed some light on the debate, new ways to view quantum mechanics are very welcome. The aim of this paper is to present quantum mechanics in new geometrical terms, as part of a larger set of dynamical theories, which are characterized by maps between two Riemmanian spaces, the source and the target spaces. We restrict ourselves to the case of $l$ non-relativistic particles. In this language, we were able to address one fundamental question which is not generally considered in quantum mechanics: why the wave function should live in the complex plane, and not in some other more contrived mathematical space? In our formalism, appealing to the geometrical quantities that we have at our disposal, we were able to give arguments on why the target space should be 2-dimensional.

At first, we review the theory of maps between a source manifold $\textbf{M}$ and a target manifold $\textbf{N}$. Then, restricting ourselves to map equations involving only first order time derivatives, we define the most general time-dependent map equation allowing a  conservation law, implying that the target space should be $2$-dimensional. All terms in the map equation have well defined geometrical meanings, one being the Hodge dual of the tension, and the other being proportional to one Killing vector of $\textbf{N}$. Afterwards, we make the connection of this mathematical setting to Physics. We identify the source manifold with configuration space, and we impose that the kernel of the map are the regions in configuration space where the physical system cannot be, due to boundary conditions and/or special initial-final conditions. Then, identifying an arbitrary scalar appearing in the map equation as the potential energy of the physical system, the arbitrary constants present in it are shown to have dimensions of action. The map equation can be projected in the two orthogonal directions of the target space. The first one yields the conservation law equation, and the associated current is the pull-back of the rotational Killing one-form on the target space. This conservation law yields a probability interpretation for one of the geometrical quantities defined in $\textbf{N}$. The current can be used to construct a velocity field in the source space. In this way, a general theory of motion for the $l$ non-relativistic particles was obtained, with two very special cases: classical mechanics for individual systems, and quantum mechanics. The exceptionality of these two cases among all other possibilities is described in detail, yielding a physical meaning for the target space $\textbf{N}$.

With the geometrical description of this generalized mechanics, new insights on the peculiarities of quantum mechanics arise, with immediate possible generalizations. We hope that our construction can pave the way for new fruitful developments.

The paper is divided as follows: in section II we summarize the theory of manifold mappings we will use. In section III we apply the mathematical setup to quantum physics, and in section IV we present the general theory of motion in which classical and quantum mechanics are inserted. Section V deals with possible interpretation issues and we end up with the conclusions, with the many possible developments that can be explored.

\section{\label{sec:Map}Manifold Mappings}

\subsection{Geometrical Setting}
In this section, we set notation, conventions and introduce some geometrical concepts to be used throughout. To start with, we let $(\textbf{M},\ g)$, $(\textbf{N},\ h)$ denote smooth orientable manifolds with metrics $g$, $h$, respectively. The m-dimensional manifold $\textbf{M}$ is called the \textit{source space} whereas the n-dimensional manifold $\textbf{N}$ is called the \textit{target space}. If $x^{a}$ $(a=1,...,m)$ and $y^{\alpha}$ $(\alpha=1,...,n)$ denote local coordinates in open subsets $\textbf{U}\subseteq \textbf{M}$ and $\textbf{V}\subseteq \textbf{N}$, we write the metrics in these regions as:
\begin{equation}
g=g_{ab}\ dx^{a}\otimes dx^{b},\quad\quad h=h_{\alpha\beta}\ dy^{\alpha}\otimes dy^{\beta}.
\end{equation}
In general, we use lower-case latin indices for tensors associated to $\textbf{M}$ and lower-case greeks for those associated to $\textbf{N}$, whereas partial differentiation with respect to $x^{a}$ or $y^{\alpha}$ will be often abbreviated as $\partial_{a}$ and $\partial_{\alpha}$.

In what follows, we shall be concerned with families of maps of the form
\begin{equation}
 f: (\textbf{M},\ g)\rightarrow (\textbf{N},\ h).
\end{equation}
Unless otherwise specified, all maps and associated sections will be smooth, for simplicity. A map such that $f(\textbf{U})\subseteq \textbf{V}$ is given in terms of the $n$ real functions
\begin{equation}
 y^{\alpha}=f^{\alpha}(x),
\end{equation}
which prescribe how one locally `applies' the source (or a submanifold of it) onto the target. The full map may be obtained by combining such local applications until we cover the whole of $\textbf{M}$.

The \textit{differential} of $f$ at $x\in \textbf{M}$ is, in a sense, the best linear approximation of $f$ near $x$ and induces the linear map between the tangent spaces:
\begin{equation}
d_{x}f: T_{x}\textbf{M}\rightarrow T_{f(x)}\textbf{N}.
\end{equation}
In coordinates, we have:
\begin{equation}
d_{x}f=\partial_{a}f^{\alpha}\big|_{x} dx^{a}\big|_{x}\otimes\partial_{\alpha}\big|_{f(x)}.
\end{equation}
Note that the $m\times n$ matrix of first-order partial derivatives $\partial_{a}f^{\alpha}$ is nothing but the Jacobian matrix of the map and can be used to pullback or pushforward different kinds of tensors. Formally, $df$ is a section of the tensor product bundle
\begin{equation}
E:= T^{*}\textbf{M}\otimes f^{-1}T\textbf{N},
\end{equation}
i.e. it is a tensor field of mixed type which can be thought as a $f^{-1}T\textbf{N}$-valued 1-form living in the source space.

The Riemannian structures associated to $\textbf{M}$ and $\textbf{N}$ naturally induce a covariant differentiation $\mathfrak{D}$ in $E$. In particular, the quantity $\mathfrak{D}df$ (also called the \textit{second fundamental form}) is given by
\begin{equation}\label{covariant}
\mathfrak{D}_{a}\partial_{b}f^{\alpha}:=\partial_{a}\partial_{b}f^{\alpha}-^{M}\Gamma^{c}_{\phantom a ab}\partial_{c}f^{\alpha}+^{N}\Gamma^{\alpha}_{\phantom a \beta\gamma}\partial_{a}f^{\beta}\partial_{b}f^{\gamma}
\end{equation}
where $^{M}\Gamma^{c}_{\phantom a ab}$ and $^{N}\Gamma^{\alpha}_{\phantom a \beta\gamma}$ are the Christoffel symbols corresponding to the metrics $g$ and $h$, respectively. Generalization to objects with more than one internal index follows the same route: simply add one connection $^{N}\Gamma^{\alpha}_{\phantom a \beta\gamma}$ term to `covariantise' each of the internal indices. In particular, there follow:
\begin{equation}
 \mathfrak{D}_{a}g_{bc}=0\quad\quad\phantom a \mathfrak{D}_{a}h_{\alpha\beta}=0,
\end{equation}
which means that contractions with metrics commute with covariant derivations. It is worth noticing that expression (\ref{covariant}) is bi-covariant in the sense that its geometrical meaning do not depend on coordinate re-parametrizations both in $\textbf{M}$ or $\textbf{N}$. Note also that
\begin{equation}\label{formula}
\mathfrak{D}_{a}T^{\alpha_{1}...\alpha_{p}}_{\quad\quad\phantom a\beta_{1}...\beta_{q}}=T^{\alpha_{1}...\alpha_{p}}_{\quad\quad\phantom a\beta_{1}...\beta_{q}||\lambda}\partial_{a}f^{\lambda}
\end{equation}
with $||$ the usual covariant derivative compatible with $h$.

The trace of the second fundamental form with respect to $g$ is denoted by $\tau(f)$ and is called the tension field for the map
\begin{equation}
\tau(f):={\rm trace}_{g}\mathfrak{D} df
\end{equation}
implying that $\tau\in f^{-1}TN$. In other words, one can think in $\tau$ as a rule assigning to the point $x$ a vector living in the tangent space of $\textbf{N}$ at the point $f(x)$.
More explicitly, we have:
\begin{equation}
\label{tension}
\tau^{\alpha}=\frac{1}{\sqrt{g}}\partial_{a}\left(\sqrt{g}g^{ab}\partial_{b}f^{\alpha}\right)+{}^{N}\Gamma^{\alpha}_{\phantom a \beta\gamma}\partial_{a}f^{\beta}\partial_{b}f^{\gamma}g^{ab},
\end{equation}
where the first term in he RHS is nothing but the Laplace-Beltrami operator $\nabla^{2}_{g}$ acting on $f^{\alpha}$. A map with zero tension is called a \textit{harmonic map} if $\textbf{M}$ is Riemmanian and a \textit{wave map} when $\textbf{M}$ is pseudo-Riemmanian (in what follows we deal only with Riemmanian manifolds). Important examples of these maps appear in the study of geodesics, minimal surfaces, harmonic functions, membranes, Skyrme-like-models among others. The theory of harmonic maps was first established by Eells and Sampson \cite{Eells}, while interesting applications to physics are found in Misner \cite{Misner} and Wong \cite{Wong} (see also \cite{Bruhat}).

\subsection{Schr\"odinger maps}

From now on, we will only consider 2-dimensional targets. To be specific, we assume that $(\textbf{N},\ h)$ is orientable, simply connected and rotationally symmetric about a \textit{base point} $\mathcal{O}$. In the end of this section, we will discuss the reasons for this class choice of target spaces. These targets carry a one-parameter family of rotational isometries and admit a local chart, centred at $\mathcal{O}$, such that the metric reads
\begin{equation}\label{metric}
h=dR\otimes dR+G^{2}(R)d\Phi\otimes d\Phi,
\end{equation}
with $G(0)=0$. Well known examples include the maximally symmetric cases, surfaces of revolution embedded in three-space and \textit{squashed} surfaces along the $z$ axis. We shall see next that, for these spaces, there is a natural generalization of the Schr\"odinger equation leading automatically to a conservation law.

An important ingredient in our construction will be the \textit{area function} $\mathcal{A}\in\mathcal{G}(\textbf{N})$ defined by
\begin{equation}
\mathcal{A}(R):=2\pi\int_{0}^{R} G(R')dR'.
\end{equation}
Essentially, this function is defined as the area of a disk with geodesic radius $R$ and centered in $\mathcal{O}$. It is clear that its existence is an intrinsic property of the manifold that holds regardless of whether we even employ coordinates, but with a precise physical meaning as it indicates how far is the boundary of the disk from the null map associated with unaccessible physical regions. Interestingly, the \textit{covariant Hessian} of $\mathcal{A}$ is given by
\begin{equation}\label{hess}
\mathcal{A}_{||\alpha||\beta}=e^{2\sigma}h_{\alpha\beta}
\end{equation}
with $\sigma$ a function of position (whose explicit form is unnecessary for our discussion here). Also, the area function naturally induces the one-form $K\in\Omega^{1}(\textbf{N})$, defined by:
\begin{equation}
K=\star(d\mathcal{A})/2\pi
\end{equation}
with $\star$ denoting Hodge dualization with respect to $h$ and the factor $(2\pi)^{-1}$ introduced for convenience. Hence, one obtains the vector field
\begin{equation}\label{Kill}
K^{\alpha}:=(\epsilon^{\alpha\beta}\partial_{\beta}\mathcal{A})/2\pi
\end{equation}
with $\epsilon_{\alpha\beta}$ the Levi-Civita tensor in $\textbf{N}$. A direct calculation using (\ref{hess}) reveals that $K^{\alpha}$ is a Killing vector, i.e. $\mathcal{L}_{K}h=0$. In components, we have
\begin{equation}
\label{killing-eq}
K_{\alpha||\beta}+K_{\beta||\alpha}=0.
\end{equation}
In other words, $K^{\alpha}$ is the generator of infinitesimal rotations about the base point $\mathcal{O}$. With respect to the coordinate patch (\ref{metric}), $K^{\alpha}$ is nothing but the azymuthal vector $-\partial/\partial\Phi$.

Let $f_{t}$ be a one-parameter family of smooth maps
\begin{equation}
f_{t}:(\textbf{M},\ g)\rightarrow (\textbf{N},\ h),
\end{equation}
represented in coordinates by $y^{\alpha}=f^{\alpha}(t,x)$. The image of a point $x\in\textbf{M}$ is a trajectory in $\textbf{N}$, whose tangent is expressed by the quantity $\partial_{t}f^{\alpha}\in T_{f(t,x)}\textbf{N}$. Our goal is to construct a bi-covariant map equation where a conservation law emerges, and in which the associated density might be interpreted as a probability density. The natural quantities at our disposal also belonging to $T_{f(t,x)}\textbf{N}$ as $\partial_{t}f^{\alpha}$ are the tension $\tau^{\alpha}$, defined in Eq.~\eqref{tension}, and the Killing vector field $K^{\alpha}$, defined in Eq.~\eqref{Kill}. Hence, the most general map equation with the above desired properties is what we shall call a \textit{Schr\"odinger map}. This is a family $f_{t}$ such that
\begin{equation}\label{alt}
\partial_{t}f^{\alpha}=c_{1}\star\tau^{\alpha}(f)+c_{2}FK^{\alpha}(f),
\end{equation}
where $c_1$ and $c_2$ are real constants to be determined later on, $\star\tau$ is the Hodge dual of the tension and $F$ is an arbitrary bi-covariant scalar which might depend on $t$, $x$ and the map $f$.

Given the particular differential operators involved, equation (\ref{alt}) constitute a system of second order nonlinear evolution equations for the map. The system relates well defined geometrical quantities in the tangent spaces of $(\textbf{N}, h)$ and is manifestly bi-covariant by construction. It is important to keep in mind, however, that Schr\"odinger maps will be, in general, many to one and we may need to deal with several different tensions at the same point $f(t,x)$.

It is obvious that we can compose $f_{t}$ with $\mathcal{A}$ to construct a map $(\mathcal{A}\circ f_{t}): \mathbb{R}\times\textbf{M}\rightarrow\mathbb{R}$. It turns out that if $f_{t}$ is a solution of (\ref{alt}), the quantity $\mathcal{A}(f^{\alpha}(t,x))$ satisfies the conservation law
\begin{equation}\label{conserv}
\partial_{t}\mathcal{A}+\mathfrak{D}^{a}J_{a}=0,
\end{equation}
with the \textit{current co-vector} given by
\begin{equation}\label{curr}
J_{a}(t,x):=2\pi c_{1} K_{\lambda}\partial_{a}f^{\lambda}.
\end{equation}
From this definition, it is clear that the current is proportional to the pull-back along the map of the Killing one-form, i.e. $J\propto f^{*}K$. We will normally think in $J$ as a one parameter family of one-form fields in $\textbf{M}$.

The proof of (\ref{conserv}) is straightforward and is based on the following identity
\begin{equation}
\partial_{t}\mathcal{A}=\partial_{\alpha}\mathcal{A}\left(c_{1}\star\tau^{\alpha}+c_{2}FK^{\alpha}\right),
\end{equation}
where we have used the chain rule and Eq.~(\ref{alt}). As $K^{\alpha}\partial_{\alpha}\mathcal{A}=0$, by construction, we have after some simple manipulations
\begin{equation}\label{kappatau}
\partial_{t}\mathcal{A}=-2\pi c_{1}K_{\alpha}\tau^{\alpha}.
\end{equation}
Recalling that the tension is given by $\tau^{\alpha}=\mathfrak{D}^{a}\partial_{a}f^{\alpha}$ and applying the Leibniz rule, we obtain
\begin{equation}
\partial_{t}\mathcal{A}=2\pi c_{1}[(\mathfrak{D}^{a}K_{\lambda})\partial_{a}f^{\lambda}-\mathfrak{D}^{a}(K_{\lambda}\partial_{a}f^{\lambda})].
\end{equation}
Now, using formulas (\ref{formula}) and (\ref{killing-eq}) one concludes that the first term in the RHS identically vanishes, implying in (\ref{conserv}) with (\ref{curr}).

A remarkable consequence of the conservation law (\ref{conserv}) is that it can be interpreted in a fully geometrical fashion. In order to conclude this, recall that the graph of a map $f:\textbf{M}\rightarrow\textbf{N}$ is the subset of $\textbf{M}\times\textbf{N}$ defined by
\begin{equation}
\mbox{graph}(f)=\{(x,f(x))|x\in\textbf{M}\},
\end{equation}
and that, since $T(\textbf{M}\times\textbf{N})\cong T\textbf{M}\times T\textbf{N}$, we can define a Riemannian metric in the product as follows:
\begin{eqnarray}
\label{bigG}
&&g^{\textbf{M}\times\textbf{N}}_{(x,y)}:=T_{(x,y)}(\textbf{M}\times\textbf{N})\times T_{(x,y)}(\textbf{M}\times\textbf{N})\rightarrow \mathbb{R}\nonumber\\
&&((v_{1},w_{1}),(v_{2},w_{2}))\mapsto g_{x}(v_{1},v_{2})+h_{y}(w_{1},w_{2}).
\end{eqnarray}
Now, the image of a point $x\in\textbf{M}$ by a Schr\"odinger map defines a disk $\mathcal{D}_{t}(x)\subseteq\textbf{N}$, whose center is $\mathcal{O}$ and area $\mathcal{A}(f_{t}(x))$. Therefore, $f_{t}:\textbf{M}\rightarrow\textbf{N}$ induces a stack of disks in $\textbf{M}\times\textbf{N}$, for all times (the graph is tangent to the stack). The volume of such (m+2)-dimensional solid with respect to $g^{\textbf{M}\times\textbf{N}}$, which might depend on the time parameter, is given by
\begin{equation}
\label{const}
\mathcal{V_T}:=\int_{\textbf{M}}\mathcal{A}(f_{t}(x))dv_{g},
\end{equation}
with $dv_{g}$ being the element of volume associated to $g$. Considering only the subspace of maps with finite $\mathcal{V_T}$, and mild assumptions on the behavior of $\partial_{k}f$ (namely, when the source space $\textbf{M}$ is non-compact, we shall assume that $f(t,\infty)=\mathcal{O}$, $\forall t$), we obtain that $d\mathcal{V_T}/dt=0.$ In other words, the continuity equation (\ref{conserv}) implies, under these conditions, in the conservation of a volume in $\textbf{M}\times\textbf{N}$.

Reviewing what we have achieved in this subsection, equation~\eqref{alt} is the most general first order in time map equation where the conservation law \eqref{conserv} is satisfied, which makes the Schr\"odinger map very special. This comes, as we have seen, from the geometrical properties of $\epsilon^{\alpha}_{\phantom a\beta}\tau^{\beta}$, and the fact that the Killing field $K_{\alpha}$ is orthogonal to the gradient of area. This property allows the introduction of the arbitrary term $F(t,x,f)K^{\alpha}$ in the RHS of Eq.~\eqref{alt} without loss of this fundamental property. Of course, one can also define a \textit{free Schr\"odinger map} for the particular case when $F(t,x,f)=0$.

We can now discuss the reasons for choosing a $2$-dimensional target space with rotational symmetry. As we will see in the next section, we will restrict ourselves to maps which are \textit{null} when applied to regions in the source configuration space $(\textbf{M},\ g)$ that are unaccessible to the physical system, due to either boundary conditions or initial-final conditions, hence distinguishing the origin $\mathcal{O}$ of the target space $(\textbf{N},\ h)$, called the base point, around which a rotational symmetry can be naturally defined. The Hodge dual in $\epsilon^{\alpha}_{\phantom a\beta}\tau^{\beta}$ is crucial for the existence of the conservation law, and a consequence of its presence is the possibility of describing a flat 2-dimensional target as the usual complex plane of quantum mechanics. If one would like to enlarge the dimensionality of the target space, the Hodge dual of the tension would bring about tensors with rank bigger than one, which cannot be directly associated with the vector $\partial_{t}f^{\alpha}$. In order to make this association, further derivatives of the tension and their combinations must be used, and the map equation would not be anymore a second order equation for the map. Finally, the rotational symmetry of the $2$-dimensional target space with its associated Killing vector $K^{\alpha}$ allows the introduction of the term $F(t,x,f)K^{\alpha}$. Without this target symmetry, such generalization would not be possible.

\section{\label{sec:Phys}Application to Quantum Physics}

In the previous section we presented in general terms the class of maps we are interested in. Let us now apply this mathematical framework to the quantum theory of $l$ non-relativistic spinless particles. In this case, the relevant degrees of freedom are points in a $3l$ dimensional configuration space (we are considering a $3-$dimensional physical space, but the description can be straightforwardly generalized to any number of physical space dimensions), or surfaces on it, as in the case of holonomous constrained systems. We have thus two spaces: the $m=3l$-dimensional configuration space $\textbf{M}$ containing the degrees of freedom of the physical system, and the $n=2$-dimensional space $\textbf{N}$. We will consider $\textbf{N}$ to be endowed with a flat Riemmanian geometry. In the Conclusion we will discuss the consequences of relaxing this assumption. When the target manifold coincides with the Euclidean plane, equation (\ref{alt}) reduces to the usual Schr\"odinger equation with appropriate choices of the constants $c_{1}$, $c_{2},$ and the arbitrary function $F$. Within this framework, if one interprets the area function $\mathcal{A}(f_{t})$ as the probability density distribution and $J_{a}$ as the probability current, Eq. (\ref{conserv}) implies in the unitary evolution of the wave function, as required by ordinary quantum mechanics. It should be stressed, however, that the approach presented up to now do not rely on any particular type of interpretation chosen for quantum mechanics. Rather it sheds light into the geometrical aspects of the theory while keeping its interpretation arbitrary.

Henceforth, to obtain ordinary quantum mechanics, we restrict ourselves to the one-parameter family of maps
\begin{equation}
f_{t}:(\textbf{M},g)\rightarrow(\mathbb{R}^{2},h)
\end{equation}
In order to connect Eq.~\eqref{alt} to the usual Schr\"odinger equation, the quantity $F$ must contain the classical potential energy, hence it should have dimension of energy, $[F]=E$ (the brackets mean``the physical dimension of").  The coordinates of the source space have dimensions of length, $[x^a]=L$, and the metrics $g_{ab}$ and $h_{\alpha\beta}$ are dimensionless. If one wants to normalize the constant ${\mathcal{V}}$ in Eq.~\eqref{const} to unity, then $[f^\alpha]=L^{-m/2}$. As a consequence, we obtain from Eq.~\eqref{alt} that $[c_1]=L^2/T=ET/M$ and $[c_2]=(ET)^{-1}$. Hence we just define $\hbar := 1/c_{2}$ and $c_1=:-\hbar/(2m_0)$, the constant $m_0$ being a reference mass parameter. The masses of the particles may be properly inserted in the components of $g^{ab}$ through the dimensionless parameters $m_0/m_i$, $m_i$ being the mass of the $i$-th particle. In the evolution equations, the parameter $m_0$ always appears together with $g^{ab}$. In this way, we can absorb the masses of the particles in $g^{ab}$, still keeping it dimensionless.

With these conventions, Eq.~\eqref{alt} becomes
\begin{equation}\label{alt2}
\partial_{t}f^{\alpha}=-\frac{\hbar}{2m_{0}}\star\tau^{\alpha}+\frac{1}{\hbar}FK^{\alpha}
\end{equation}
and the current co-vector reads
\begin{equation}
J_{a}=-\frac{\pi\hbar}{m_{0}}K_{\alpha}\partial_{a}f^{\alpha}.
\end{equation}

In the case that $\textbf{N}$ is flat, there are some useful identities:
\begin{equation}
\label{hessflat}
\mathcal{A}_{||\alpha||\beta}=2\pi h_{\alpha\beta}  ,
\end{equation}
\begin{equation}
\label{normareagrad}
h^{\alpha\beta}\partial_{\alpha}\mathcal{A}\partial_{\beta}\mathcal{A}=4\pi\mathcal{A}  ,
\end{equation}
\begin{equation}
\label{ke}
K_{\alpha||\beta}=\epsilon_{\alpha\beta}  ,
\end{equation}
\begin{equation}
\label{knorm}
K^{\alpha}K_{\alpha}=\frac{\mathcal{A}}{\pi}  ,
\end{equation}
\begin{equation}
\label{kAe}
K_{\alpha}\partial_{\beta}\mathcal{A}-K_{\beta}\partial_{\alpha}\mathcal{A}=2\mathcal{A}\epsilon_{\alpha\beta}.
\end{equation}

An interesting byproduct of the bi-covariant approach presented thus far is the following. The formalism suggests that there is a special role played by the radial and angular directions in the target space. We may arrive at this conclusion reasoning in purely geometrical terms, without evoking specific coordinates in $\textbf{N}$, or specifying the function $F$.
Indeed, it can be checked that Equation \eqref{conserv} can be understood as the projection of Eq.~\eqref{alt2} in the direction of $\partial_{\alpha} \mathcal{A}$.
The other independent equation coming from Eq.~\eqref{alt2} can be obtained through its projection in the orthogonal direction to the gradient of the area, which is the direction of the Killing vector $K^{\alpha}$. This projection reads
\begin{equation}
\label{Kproj}
K_{\alpha}\partial_{t}f^{\alpha}=-\frac{\hbar}{4\pi m_{0}}\tau^{\alpha}\partial_{\alpha}\mathcal{A}+\frac{F}{\pi\hbar}\mathcal{A}
\end{equation}
where the last term is a consequence of the identity \eqref{normareagrad}.

Note that the term involving the tension splits in two terms with precise geometrical meanings
\begin{equation}\label{taugrad}
\tau^{\alpha}\partial_{\alpha}\mathcal{A}=\mathfrak{D}^{a}(\partial_{a} f^{\alpha}\partial_{\alpha}\mathcal{A})-2\pi h_{\alpha\beta}\partial^{a}f^{\alpha}\partial_{a}f^{\beta}.
\end{equation}
We recognize the first term on the RHS as the divergence of $f^{*}d\mathcal{A}$ and the second as the Dirichlet energy for the map (see \cite{Wong} for details).
There is another suggestive form for Eq. \eqref{taugrad}. Indeed, defining the scalar
\begin{equation}
\label{qp}
Q=-\frac{\hbar^{2}}{2m_{0}}\frac{\nabla^{2}_g\mathcal{A}^{1/2}}{\mathcal{A}^{1/2}},
\end{equation}
where $\nabla^{2}_g:=\mathfrak{D}^a \mathfrak{D}_a$,
the normalized vector in the direction of the gradient of the area (see Eq.~\eqref{normareagrad}),
\begin{equation}
v_\alpha := \frac{\partial_{\alpha}\mathcal{A}}{2\sqrt{\pi \mathcal{A}}},
\end{equation}
and the projector in its orthogonal direction,
\begin{equation}
P_{\alpha\beta}=h_{\alpha\beta}-v_\alpha v_\beta =h_{\alpha\beta}-\frac{\partial_{\alpha}\mathcal{A}\partial_{\beta}\mathcal{A}}{4\pi \mathcal{A}} ,
\end{equation}
one obtains
\begin{equation}\label{projected}
\hbar K_{\alpha}\partial_{t}f^{\alpha}=\frac{\hbar^2}{2m_{0}}P_{\alpha\beta}\partial_{a}f^{\alpha}\partial^{a}f^{\beta}+\frac{(Q+F)}{\pi}\mathcal{A}.
\end{equation}
Defining the other unit vector $w_\alpha$ orthogonal to $v_\alpha$ as
\begin{equation}
w_\alpha := \sqrt{\frac{\pi}{\mathcal{A}}}K_{\alpha},
\end{equation}
the metric on $\textbf{N}$ can also be written as
\begin{equation}
h_{\alpha\beta}=\frac{\pi}{\mathcal{A}}K_\alpha K_\beta +
\frac{\partial_{\alpha}\mathcal{A}\partial_{\beta}\mathcal{A}}{4\pi \mathcal{A}} ,
\end{equation}
yielding $$P_{\alpha\beta}=\frac{\pi}{\mathcal{A}}K_\alpha K_\beta.$$
Hence one gets
\begin{equation}\label{projected15}
\hbar \frac{\pi}{\mathcal{A}} K_{\alpha}\partial_{t}f^{\alpha}=
\frac{\hbar^2}{2m_{0}}\frac{\pi^2}{{\mathcal{A}}^2}g^{ab}K_{\alpha}\partial_{a}f^{\alpha}K_{\beta}\partial_{b}f^{\beta}+Q+F.
\end{equation}
This projected equation must always go along with the projected equation leading to the conservation law, namely
\begin{equation}
\label{cons2}
\partial_{t}\mathcal{A}+\mathfrak{D}^{a}J_{a}=0,
\end{equation}
with
\begin{equation}
\label{curr2}
J_{a}(t,x):=-\frac{\pi\hbar}{m_0} K_{\lambda}\partial_{a}f^{\lambda}.
\end{equation}

We can put these three equations in a yet simpler form. Using equations (\ref{hessflat}), (\ref{normareagrad}), (\ref{ke}), (\ref{knorm}) and (\ref{kAe}), we can show that,
\begin{equation}
\partial_{b}\biggl(\frac{K_{\alpha}\partial_{a}f^{\alpha}}{\mathcal{A}}\biggr) = \partial_{a}\biggl(\frac{K_{\alpha}\partial_{b}f^{\alpha}}{\mathcal{A}}\biggr),
\end{equation}
\begin{equation}
\partial_{a} \biggl(\frac{K_{\alpha}\partial_{t}f^{\alpha}}{\mathcal{A}}\biggr) = \partial_{t}\biggl(\frac{K_{\alpha}\partial_{a}f^{\alpha}}{\mathcal{A}}\biggr).
\end{equation}
The first equation implies that we can write $K_{\alpha}\partial_{a}f^{\alpha}/\mathcal{A}$ as a gradient. Hence we define,
\begin{equation}\label{Sdef}
\frac{\hbar\pi K_{\alpha}\partial_{a}f^{\alpha}}{\mathcal{A}} \equiv -\partial_{a} S .
\end{equation}
Fom the second equation we obtain that
\begin{equation}
\frac{\hbar\pi K_{\alpha}\partial_{t}f^{\alpha}}{\mathcal{A}} = -\partial_{t} S + g(t),
\end{equation}
where $g(t)$ is an arbitrary function of $t$. However, as we will see in the sequel, writting the above equations in polar coordinates we can easily show that $g(t)=0$.
Hence, the three equations \eqref{projected}, \eqref{cons2}, \eqref{curr2}, can be written as:
\begin{equation}\label{projected2}
\partial_{t}S +
\frac{1}{2m_{0}}g^{ab}\partial_{a}S \partial_{b}S+Q+F=0,
\end{equation}
and
\begin{equation}
\label{cons3}
\partial_{t}\mathcal{A}+\mathfrak{D}^{a}J_{a}=0,
\end{equation}
with
\begin{equation}
\label{curr2}
J_{a}(t,x):=\frac{\mathcal{A}}{m_0} \partial_{a}S.
\end{equation}
Note that $\hbar$ disappeared from these equations, except inside $Q$ (assuming that $F$ does not depend on $\hbar$), which inevitably arises from the tension $\tau^\alpha$. This is an important remark, which will be discussed in the following sections.

Let us now write these equations in the two coordinates suitable to the problem: Cartesian and polar coordinates of flat $\textbf{N}$ space.

%In this last case, one can define the two bases functions $e_1^\alpha :=v^\alpha$ and $e_2^\alpha :=K^\alpha$, yielding $h_{11}=e_1^\alpha e_1^\beta h_{\alpha\beta}=1$, $h_{22}=e_2^\alpha e_2^\beta h_{\alpha\beta}=\mathcal{A}/\pi$, and $h_{12}=e_1^\alpha e_2^\beta h_{\alpha\beta}=0$. As $\mathcal{A}$ is positive definite, one can write $\mathcal{A}/\pi =: R^2$, which, together with the angular coordinate $\Phi$ associated to the rotational Killing vector $K^\alpha$, leads us to polar coordinates.

\subsection{Cartesian coordinates}
As we are assuming that the target space $\textbf{N}$ is flat, we can choose Cartesian coordinates $(X_{1},X_{2})$, such that $h={\rm {diag}}(1,1)$. The area function and the Killing vector read as
\begin{equation}
\mathcal{A}=\pi(X_{1}^{2}+X_{2}^{2})\quad\quad K^{\alpha}=(X_{2}, -X_{1})
\end{equation}
As the connections ${}^{N}\Gamma$ all vanish, we obtain
\begin{equation}\nonumber
\star\tau^{1}=\nabla^{2}_g X_{2},\quad\quad\star\tau^{2}=-\nabla^{2}_g X_{1}.
\end{equation}
Substitution in (\ref{alt2}) gives the linear equations
\begin{eqnarray}
\label{cart1}
\partial_{t}X_{1}&=&-\frac{\hbar}{2m_0}\nabla^{2}_g X_{2}+\frac{F}{\hbar}X_{2}\\
\label{cart2}
\partial_{t}X_{2}&=&+\frac{\hbar}{2m_0}\nabla^{2}_g X_{1}-\frac{F}{\hbar}X_{1}
\end{eqnarray}
whereas the current has the form
\begin{equation}
J_{a}=\pi\frac{\hbar}{m_0}(X_{1}\partial_{a}X_{2}-X_{2}\partial_{a}X_{1}).
\end{equation}
Defining the complex quantity $\Psi = X_1 + i  X_2$ and choosing $F=V(x,t)$, we can recast equations \eqref{cart1} and \eqref{cart2} in the compact form
\begin{equation}
\label{sch}
i\hbar \partial_{t}\Psi=-\frac{\hbar^2}{2m_0}\nabla^{2}_g \Psi+V\Psi,
\end{equation}
which is nothing but Schr\"odinger equation.
\subsection{Polar Coordinates}
We now have $y^{\alpha}=(R,\Phi)$ with $h={\rm {diag}}(1, R^{2})$. The area function and the Killing vector have the form
\begin{equation}
\label{area-polar}
\mathcal{A}=\pi R^{2}\quad\quad K^{\alpha}=(0, -1)
\end{equation}
while the non-vanishing connections are
\begin{equation}
 \Gamma^{R}_{\phantom a\Phi\Phi}=-R\quad\quad\Gamma^{\Phi}_{\phantom a R\Phi}=R^{-1}
\end{equation}
A direct calculation yields
\begin{eqnarray*}
\star\tau^{1}&=&R\nabla^{2}_g \Phi+2\nabla R.\nabla\Phi\\
\star\tau^{2}&=&\nabla\Phi.\nabla\Phi-\nabla^{2}_{g}R/R,
\end{eqnarray*}
where $\nabla f^\alpha.\nabla f^\beta := \partial_a f^\alpha \partial_b f^\beta g^{ab}$. The components of Equation (\ref{alt2}) take the form
\begin{eqnarray}
\label{dbb0}
\partial_{t}R&=&-\frac{\hbar}{2m_0}(R\nabla^{2}_g\Phi+2\nabla R . \nabla\Phi)\nonumber\\
\partial_{t}\Phi&=&\frac{\hbar}{2m_0}\left(\frac{\nabla^{2}_g R}{R}-\nabla\Phi.\nabla\Phi\right)-\frac{F}{\hbar}.
\end{eqnarray}
Alternatively, defining the action function $S=\hbar\Phi$ and choosing $F=V(x,t)$ one can rewrite Eqs.~\eqref{dbb0} as
\begin{eqnarray}
\label{dbbR}
&&\partial_{t}R^2 + \frac{\nabla}{m_0}. (R^2\nabla S)=0\\
\label{dbbS}
&&\partial_{t}S + \frac{\nabla S.\nabla S}{2m_0} + V - \frac{\hbar^2}{2m_0}\frac{\nabla^{2}_g R}{R}=0.
\end{eqnarray}
which are nothing but the coupled equations which emerge when we write $\Psi=Re^{iS/\hbar}$ . Also, the current reads as
\begin{equation}
\label{current-polar}
J_{a}=\frac{\pi}{m_0} R^{2}\partial_{a} S.
\end{equation}
Comparing Eq.~\eqref{dbbS} with Eq.~\eqref{projected2}, one can see that the arbitrary function $g(t)$ must indeed be null.

\section{\label{sec:Motion}General theory of motion}

%As we have argued in the previous section, the $2$-dimensional $\textbf{N}$ space yields a $2$-fold information about the physical system under consideration: its law of motion and their unavoidable uncertainties.

Looking at Eqs.~\eqref{conserv} and \eqref{const}, it is natural to interpret the area function in $\textbf{N}$ as a probability density distribution.
%, furnishing information about the uncertainties in the physical system. By its definition, the area function is completely determined at each point in $\textbf{N}$ through the radial coordinate of this point with respect to $\mathcal{O}$.
Its gradient defines one of the two orthogonal directions in $\textbf{N}$. The other orthogonal independent direction is given by the rotational Killing vector $K_{\alpha}$. Its pull-back into $\textbf{M}$ defines the current $J_a$. Also, $J^a/\mathcal{A} = g^{ab}J_b/\mathcal{A}$ is a vector in $M$ with velocity dimensions, $[J^a/\mathcal{A}]=L/T$. Hence, in the same way we have established a connection between the tangent vector in $\textbf{N}$, $\partial_t f^{\alpha}$, with geometrical quantities in it (see Eq.~\eqref{alt2}), it is reasonable to postulate a connection between the tangent vector in $\textbf{M}$, $\dot{x}^a$, with geometrical quantities in it through the equation
\begin{equation}
\label{guidance}
\dot{x}^a = \frac{J^a}{\mathcal{A}},
\end{equation}
where the dot means time derivative. Looking at Eq.~\eqref{curr2}, one can write Eq.~\eqref{guidance} in the suggestive form:
\begin{equation}
\label{guidance2}
\dot{x}^a = \frac{g^{ab}\partial_b S}{m_0},
\end{equation}
This equation is the natural covariant generalization of the time evolution equation furnished by the Hamilton-Jacobi theory, which describes motion in configuration space. It is called the guidance equation.

%, which can be explicitly obtained from Eq.~\eqref{guidance} when one writes it using polar coordinates, see Eqs.~\eqref{current-polar} and \eqref{area-polar}.

In this language, the emergence of different physical theories will depend on the choice of $F(t,x,f)$. Let us now discuss two very special cases.

\subsection{First order classical mechanics}

Looking at Eq.~\eqref{projected2}, one can make a particular choice of $F(t,x,f)$ which render this equation with a very special form. This choice is

\begin{equation}
\label{Fclassical}
F(t,x,f) = V(t,x) -Q.
\end{equation}

In this case, not only the constant $\hbar$ disappears completely from both equations \eqref{cons3} and \eqref{projected2}, but also Eq.~\eqref{projected2} becomes independent of $\mathcal{A}$ because $Q$ disappears. Assuming that $V(t,x)$ is the potential energy of classical mechanics, Eq.~\eqref{projected2} becomes the usual Hamilton-Jacobi equation for $S$ which, together with Eq.~\eqref{guidance}, yields the classical dynamical laws governing the system of $l$ non-relativistic particles. However, this formulation is not entirely equivalent to classical mechanics. First note that $S$ is an angular coordinate, as it is associated with the rotation Killing vector $K_\alpha$, subjected to periodicity conditions, which is not the case of classical mechanics. Second, it comes from a solution of the map equation Eq.~\eqref{alt2}, and as such it depends only on $(x,t)$, $S=S(x,t)$. Hence, together with Eq.~\eqref{guidance}, it yields a single valued velocity field at each $(x,t)$, each trajectory of the many-particle system depending only on the arbitrary initial position $x_0$. This is not the case of classical mechanics, where, in the case of an ensemble of many-particle systems subjected to the same classical potential $V(t,x)$ but with different initial conditions, the $S$ coming from the Hamilton-Jacobi formulation also depends on a label distinguishing each many-particle system in the ensemble with different initial conditions, $S=S(x,t;C_i)$, and, as such, the velocities coming from it are not single valued at each $(x,t)$. In other words, in the above formulation the many-particle trajectories cannot cross in $(x,t)$ space, but in usual classical mechanics they can. This is because ordinary classical mechanics is based on second order differential equations, one also needs to prescribe the initial momenta, while in the present formulation the dynamics relies on the first order differential equation \eqref{guidance}. Only in $(x,p,t)$ space ($p$ denoting momenta) trajectories do not cross in ordinary classical mechanics. As a consequence, an initial condition distribution should be prescribed only in phase space. For a good discussion on these issues, see Ref.~\cite{classicalSch}.
Concluding, the choice \eqref{Fclassical} yields the simplest possible dynamics one can obtain from the combination of Eqs.~\eqref{alt2} and \eqref{guidance}, leading to classical mechanics for individual many-particle trajectories in configuration space. However, Eq.~\eqref{alt2} is highly non linear and complicated, and their projections, Eqs~\eqref{curr2} and \eqref{projected2}, cannot describe appropriately ensembles of classical many-particle systems, essentially because it is a first order mechanics, contrary to the second order ordinary classical mechanics. For details see Ref.~\cite{classicalSch}.

\subsection{Quantum mechanics}

Another particular choice of $F(t,x,f)$ which renders Eq.~\eqref{alt2} into another very special form is

\begin{equation}
\label{Fquantum}
F(t,x,f) = V(t,x) .
\end{equation}

In this case, making use of the Cartesian coordinates presented in Subsection III.A, Eq.~\eqref{alt2} becomes linear. Taking equations \eqref{cart1} and \eqref{cart2}, and defining $\Psi = X_1 + i  X_2$, we can rewrite these two equations in the compact form

\begin{equation}
\label{sch}
i\hbar \partial_{t}\Psi=-\frac{\hbar^2}{2m_0}\nabla^{2}_g \Psi+V\Psi.
\end{equation}

This is the only case in which the map equation can be put in linear form. Again, assuming that $V(t,x)$ is the potential energy of classical mechanics, Eq.~\eqref{alt2} reduces to the usual Schr\"odinger equation which, together with Eq.~\eqref{guidance}, yields the de Broglie-Bohm description of $l$ non-relativistic quantum particles. Now Eq.~\eqref{projected2} is not anymore independent of the area function $\mathcal{A}$, and $\hbar$ cannot be removed. The action function $S$ giving the laws of motion is now entangled through $\hbar$ in Eq.~\eqref{projected2} with the area function $\mathcal{A}$ through the quantum potential $Q$. The constant $\hbar$ gives the strength of this connection. This turns quantum mechanics fundamentally different from classical mechanics.

Contrary to the classical mechanics case, the choice \eqref{Fquantum} yields a complicated dynamics for the $l$ non-relativistic particles, but the map equation \eqref{alt2} is the simplest one, as it can be written in linear form. The superposition principle is now valid, with utmost importance for quantum physics \cite{dirac}. In this case, it is better to deal with the full map equation \eqref{alt} (specially in its complex form), than with their projections.

%\subsection{Other possible dynamics} As we have seen, classical and quantum mechanics are very special cases of the many possible dynamics which can arise from Eqs.~\eqref{alt2} and \eqref{guidance}. The mechanics arising from the choice \eqref{Fclassical} is the one with the simplest dynamical laws. They are completely independent from the area function, and $\hbar$ is irrelevant. We will discuss this peculiarity of classical mechanics in more detail in the next section.
%Of course other possibilities with these properties could be constructed, e.g. with $$ F(t,x,f) = V(t,x)(1 + K^{\alpha}_{|| \alpha} -Q,$$ but they are less simple. Note that we can construct other dynamics in which $\hbar$ is irrelevant, but nevertheless the area function is important for the laws of motion, like in $$ F(t,x,f) = V(t,x)\left(1+\frac{\mathfrak{D}_{a}\mathfrak{D}^{a}\mathcal{A}}{h_{\alpha\beta}\partial_{a}f^{\alpha}\partial^{a}f^{\beta}}\right) -Q.$$ It is clear that different choices are available, engendering many types of dynamics, but all of them are rather arbitrary. The unique linear possibility is that of de Broglie-Bohm quantum mechanics, but the dynamics in this case necessarily depends on the area function and $\hbar$ through $Q$.

\section{Interpretation issues}

The two equations \eqref{alt2} and \eqref{guidance} are the core of the formalism presented above. They form a symmetric framework in which tangent vectors in the source and target spaces are linked with geometrical objects defined in both spaces through bi-covariant equations. One very important consequence of the imposition of both equations \eqref{alt2} and \eqref{guidance} is the appearance of the notion of quantum equilibrium \cite{valentini1,Durr1,Durr2}. It comes from a property called equivariance, which emerges naturally from equations (\ref{alt2},\ref{guidance}). Indeed, any probability measure which coincides with the area function at some moment of time, will continue to be given by the area function at all times. This is because the area function evolves in time in the same way as any probability measure describing a physical system evolving according to the guidance equation \eqref{guidance}, as it satisfies Eq.~\eqref{cons3}. Hence, the area function is the unique time independent function of the time-dependent map function describing the physical system with this property, playing the role of the static measures used in statistical mechanics, even being time-dependent. This special property of the area function, called equivariance, allows its connection with empirical probability distributions at all times, which is not the case for any other theoretical probability measure: they could coincide only at one particular moment of time. Hence, typicality defined by this measure, called the quantum equilibrium measure, emerges naturally from the physics of the Bohmian dynamics itself (see Ref.~\cite{Durr2}, specially chapter 11). This line of reasoning naturally leads to the question whether a non-equilibrium measure, different from the area function, can exist in Nature, how fast they move to the equilibrium measure, under what conditions, and what are their physical consequences. The strategy is to impose mild distributions of unknown initial conditions for the physical system at hand, and show that equations (\ref{alt2},\ref{guidance}) generally take us, at a coarse grained level, to the result that the degrees of freedom of the physical system get distributed according to the area function after some time evolution, reaching quantum equilibrium. This is work in progress \cite{valentini2,valentini3,valentini4}. However, what is important for us here is the verification that the association of the area function with a probability measure is a consequence of the postulation of equations \eqref{alt2} and \eqref{guidance}: it is not postulated a priori, it comes as a second step in the formulation of the theory. Hence, a probability notion emerges from both Eq.~\eqref{guidance} and the conservation law arising from Eq.~\eqref{alt2}, yielding a physical interpretation for the area function.This result is true for whatever function $F$ appearing in Eq.~\eqref{alt2}, and that was the reason for seeking for a map equation which implies a conservation law. One could relax this imposition, allowing a map equation which has not a Schr\"odinger form. However, even in this case, which may be the situation in quantum gravity and quantum cosmology \cite{qg,Nelson}, one can still recover a probability notion in the end if the target space is $2-$dimensional from the beginning. In order to do that, one should be able to show that suitable collections of sub-maps, maps defined in subsets of the source space, satisfy an approximate Schr\"odinger map equation coming from the original one. For such sub-maps, a probability interpretation of the area can be recovered. They are called conditional maps, and examples of such situations have been already presented \cite{Durr1,Durr2}, even at the level of quantum cosmology \cite{tovar}. Hence, the bi-dimensionality of the target space is mandatory if one wants to obtain a simple conservation law, either from the beginning, or through these conditional maps.

One can also give another justification for a $2-$dimensional target space. First note that the pull-back current of the rotation Killing vector of the target space yields the law of motion in configuration space through Eq.~\eqref{guidance}. The existence of this rotation Killing vector field came from the imposition that the target space should be isotropic with respect to the origin, which is the map image of the regions in configuration space where the physical system cannot be, due either to boundary conditions or special initial-final conditions. A remarkable outcome of Eq.~\eqref{guidance} is that the $S$ function arising in Eq.~\eqref{Sdef} is not defined in the origin, and the law of motion Eq.~\eqref{guidance} does not hold there, as it should be. This special property of the law of motion Eq.~\eqref{guidance}, a consequence of the existence of the rotation Killing vector $K_{\alpha}$ in $\textbf{N}$, brings about the second argument in favor of a $2-$dimensional target space: indeed, in a $n-$dimensional isotropic target space one has $n(n-1)/2$ Killing vector fields, hence a unique guidance equation can only arise for $n=2$.

Can we give a physical interpretation for the $2-$dimensional target space? Note that, in the general case, the area function may not be necessarily dressed with a probability interpretation, only after conditional maps can be defined, and the general map equation is reduced to the Schr\"odinger map equation \eqref{alt2}.

As it has been shown in section III, the projected equations coming from Eq.~\eqref{alt2}, together with Eq.~\eqref{guidance}, can be put in the form,

\begin{equation}
\label{cons4}
\partial_{t}\mathcal{A}+\mathfrak{D}_{a}(\mathcal{A}\dot{x}^a)=0,
\end{equation}
\begin{equation}\label{projected3}
\partial_{t}S +
\frac{1}{2m_{0}}g^{ab}\partial_{a}S \partial_{b}S-\frac{\hbar^{2}}{2m_{0}}\frac{\nabla^{2}_g\mathcal{A}^{1/2}}{\mathcal{A}^{1/2}}+F=0,
\end{equation}
\begin{equation}
\label{guidance3}
\dot{x}^a = \frac{g^{ab}\partial_b S}{m_0},
\end{equation}
where we have used Eq.~\eqref{guidance} in Eq.~\eqref{cons3} to obtain Eq.~\eqref{cons4}.
We hence have a set of mixed equations, where the function $S$ guides the particles through Eq.~\eqref{guidance3}. As we have seen, only in the classical case, where we make the choice
\begin{equation}
\label{classical-choice}
F = V(x) + \frac{\hbar^{2}}{2m_{0}}\frac{\nabla^{2}_g\mathcal{A}^{1/2}}{\mathcal{A}^{1/2}}
\end{equation}
does Eq.~\eqref{projected3} gets independent from the area function. This has one remarkable consequence. Suppose the relevant universal physical interactions acting on the physical system are known ($V(x,t)$ is known). Then, if one takes any small region in configuration space surrounding a point $x_i$ at time $t_0$, knowledge of $\rho,\partial_a S,\partial_{a}\partial_{b}S$ around this point at $t_0$ is sufficient to obtain the future evolution of the congruence of curves emerging from this small region. However, any other choice of $F$ in which the term
$$Q=-\frac{\hbar^{2}}{2m_{0}}\frac{\nabla^{2}_g\mathcal{A}^{1/2}}{\mathcal{A}^{1/2}},$$ necessarily arising from the tension term in Eq.~\eqref{alt2}, still persists in Eq.~\eqref{projected3} and it is not negligible, complete knowledge of the area function $\mathcal{A}$ for all points in configuration space at $t_0$ (all their spatial partial derivatives) is necessary to calculate the evolution of the physical system around $x_i$. Hence, the presence of $Q$ in Eq.~\eqref{projected3}, which happens in quantum mechanics (see Eq.~\eqref{sch}), brings about the wave feature of quantum dynamics, linked to interference, contextuality, and non-locality, depending on the map. The area function connects the dynamics of the system with boundary conditions and/or special initial-final conditions imposed on the system as it yields the distance in $\textbf{N}$ space to its origin, the map image of the regions in $\textbf{M}$ where the physical system cannot be due to these conditions.

Concluding, the bi-dimensionality of the target space $\textbf{N}$ is the minimum necessary for taking into account the experimentally observed non-locality and wave aspects of quantum physical systems, which appears in their dynamics through $\hbar$, leading to all weird features characteristic of quantum mechanics. As we have discussed above, a one-dimensional $\textbf{N}$ leads to an essentially local dynamics, which is sufficient to describe the local classical world, but insufficient to describe the nonlocal features of the quantum world.

\section{Conclusion}

In this paper we constructed mathematical representations of physical theories describing $l$ non-relativistic particles in terms of time-dependent maps between the configuration space $\textbf{M}$ with metric $g$ containing the $3l$ degrees of freedom of the particles and a $n$-dimensional target space $\textbf{N}$ with metric $h$, from which the dynamics is obtained. Both metrics are Riemmanian. The kernel of the map is constituted by the regions in $\textbf{M}$ where the physical system cannot be, due to boundary conditions and/or special initial-final conditions. Hence, the origin of $\textbf{N}$, $\mathcal{O}$, is a special base point, and we restricted ourselves to target spaces which are isotropic with respect to it. If this base point can be chosen arbitrarily in $\textbf{N}$, then the target space must be maximally symmetric.

The isotropy of $\textbf{N}$ naturally implies the existence of Killing vector fields in it, and a scalar function $\mathcal{V}$ yielding the volume of the hyper-spheres surrounding the base point in $\textbf{N}$. Furthermore, the map allows the construction of another vector field in $\textbf{N}$, which is the bi-covariant map-tension $\tau^{\alpha}$. Finally, the time-derivative of the time-dependent map is a vector in the tangent space of $\textbf{N}$.

Now, the image of a point $x\in\textbf{M}$ of any time-dependent map in $\textbf{N}$ defines a hyper-sphere $\mathcal{D}_{t}(x)\subseteq\textbf{N}$, whose center is $\mathcal{O}$ and volume $\mathcal{V}(f_{t}(x))$. Therefore, $f_{t}:\textbf{M}\rightarrow\textbf{N}$ induces a stack of hyper-spheres in $\textbf{M}\times\textbf{N}$, for all times. The volume of such (3l+n)-dimensional solid with respect to the metric $g^{\textbf{M}\times\textbf{N}}$ induced by $g$ and $h$ (see Eq.~\eqref{bigG}) is given by
\begin{equation}
\label{constC}
\mathcal{V_T}:=\int_{\textbf{M}}\mathcal{V}(f_{t}(x))dv_{g},
\end{equation}
with $dv_{g}$ being the element of volume associated to $g$.
We have shown that the most general time-dependent maps satisfying bi-covariant second-order differential equations in which, under mild conditions, the total volume $\mathcal{V_T}$ is constant in time are solutions of Eq.~\eqref{alt}, which we called Schr\"odinger maps. The Schr\"odinger map equation \eqref{alt} contains all naturally defined vector fields in $\textbf{N}$ listed above, and it necessarily imposes that $\textbf{N}$ should be $2$-dimensional. If the arbitrary function $F$ has physical dimension of energy, then $1/c_2 =:\hbar$ must have dimension of action, yielding Eq.~\eqref{alt2}.

As the target space $\textbf{N}$ is $2$-dimensional, it has a unique rotational Killing vector field $K^{\alpha}$, and the volume function should be called the area function $\mathcal{A}$. This area function satisfies a conservation law in which the associated current $J^{a}$ is the pull-back in $\textbf{M}$ of the rotational Killing vector $K^{\alpha}$. Combining $J^{a}$ and $\mathcal{A}$, we can construct a vector field in $\textbf{M}$ with velocity dimensions, which can be identified with vectors belonging to the tangent vector space in $\textbf{M}$, $\dot{x}$, through Eq.~\eqref{guidance}. In this way, a dynamical law naturally emerges from the formalism.

Equations \eqref{alt2} and \eqref{guidance} lead to the notion of equivariance and quantum equilibrium, where a unique typical probability measure emerges, which is identified with the area function \cite{valentini1,Durr1,Durr2}. For a map equation without the Schr\"odinger form, which may be the situation in quantum gravity and quantum cosmology \cite{qg,Nelson}, one can still recover a probability notion in the end if one is able to show that suitable collections of sub-maps, maps defined in subsets of the source space, satisfy an approximate Schr\"odinger map equation coming from the original one. For such sub-maps, a probability interpretation of the area can be recovered.

Considering $\textbf{N}$ to be flat, classical mechanics for individual physical systems and quantum mechanics emerge in this unified description as very special cases. In classical mechanics, the projection of the map equation in the rotational Killing vector direction yields an equation for the action function appearing in the guidance equation (see Eqs.~\eqref{curr2} and \eqref{projected2} with $F=V-Q$) completely independent from the area function. The constant $\hbar$ disappears. We get a first order classical mechanics in Hamilton-Jacobi form (which is not always equivalent to ordinary classical mechanics for an ensemble of physical systems subjected to the same potential $V(x)$). Once one knows the universal laws governing the physical system ($V(x)$), some minimal information about the physical system in the neighborhood of any point in configuration space is sufficient to predict the evolution of this region for all times.

Quantum mechanics is the case with the simplest map equation, which can be put in linear form through the employment of Cartesian coordinates in $\textbf{N}$. However, the projected equations in the directions of the gradient of the area and of the rotational Killing vector are entangled, and difficult to handle. In this case, it is simpler to deal with the linear map equation itself, and use the superposition principle it allows. The dynamics now depends also on the area function through $\hbar$, see Eq.~\eqref{projected2} with $F=V$, which gives the distance to the image in $\textbf{N}$ of the kernel of the map, related to boundary conditions and/or initial-final conditions imposed on the physical system. Complete knowledge of $V(x)$ is not anymore sufficient to determine the evolution of any small region in $\textbf{M}$. One also needs complete knowledge of the area function in all points of $\textbf{M}$ at a fixed time, bringing about the wave and nonlocal aspects of quantum mechanics. Note that, in the classical case, the radial dimension of the target space is irrelevant for the dynamics, it is essentially one-dimensional. Hence, the bi-dimensionality of the target space is essential to describe the weird features of quantum mechanics, and it is the simplest target space which can accommodate the nonlocal features of Nature.

The quantum mechanics we obtained appears naturally in the de Broglie-Bohm \cite{Bohm,Holland} form. In order to get standard (Copenhaguen) quantum mechanics, Eq.~\eqref{guidance} must be eliminated. However, in the way we are presenting the subject, this would be a rather ad hoc procedure. Why Eq.~\eqref{guidance} should be present for some map equations and absent for others? Furthermore, eliminating it creates some extraneous difficulties (evocation of a collapse postulate, the measurement problem, and so on \cite{Bell}). Hence, in this view, the de Broglie-Bohm quantum dynamics appears to be more natural than standard quantum mechanics.

The geometrical view of generalized mechanics of $l$ non-relativistic particles we are presenting in this paper, which contains classical and quantum mechanics as particular cases, does not yet imply new physics. However, this rather different point of view offers natural generalizations which must be explored. What happens if we allow the $\textbf{N}$ space to be curved? In this case, global Cartesian coordinates are not possible, and the map equation is necessarily non-linear, even with the choice Eq.~\eqref{Fquantum}. Hence, we get a natural non-linear generalization of the Schr\"odinger equation. And what happens if we endow $N$ with some non-trivial topology? What is the relativistic generalization of this description? How can we describe quantum field theory and spinors in this framework? How different are the evolution of maps in $\textbf{N}$ space related with entangled states from the ones associated with separable states? And what are the consequences of allowing $\textbf{N}$ to be $d>2$-dimensional? Would it be necessary to evoke some compactification scheme in order to recover a conservation law? What sort of guidance equations will be available when there are other rotational Killing vectors at our disposal? As a simple extension of the target space used above, note that the spin degree of freedom of the non-relativistic electron can be described in this language using a target space $\textbf{N} = R^2 \times \mathcal{S}$, where $\mathcal{S}$ denotes the usual spin $1/2$ vector space. The most general map equation yielding a conservation law compatible with this structure is the non-relativistic Pauli equation. Also, a curved target space, like a sphere, would bring about a new fundamental constant $L$, the curvature scale of the sphere, the scale above which the manifold curvature could be experienced. Hence, non-linearity effects would be noticeable only when $\mathcal{A}\approx L^2$, or bigger. As no such non-linear effects were ever measured, we expect that $L$ be large (if not infinity, where we get back to linear quantum mechanics), hence only experiments involving very narrow wave functions could {\bf reveal} this type of non-linearity. This can be an experimental route to be explored.

%Finally, with the rotational Killing vector field and the area function defined in the target space $\textbf{N}$, one can construct the vector field $J^a/\mathcal{A}$ yielding the velocity field in $\textbf{M}$ through the guidance equation \eqref{guidance}. 
%That is why we called $\textbf{N}$ the causal space. Its bi-dimensionality allows a two-fold representation of causation, one linked to the universal laws of Nature, and the other with contingent causes coming from particular experimental set-ups yielding specific boundary conditions and/or special initial-final initial conditions. One can say that the mixing of these two causations in the quantum dynamics, which does not happen in classical mechanics, is the essence of its dualistic nature.

\begin{acknowledgments}
One of us, N.P.N., would like to thank CNPq of Brazil for financial support under grant 309073/2017-0.
\end{acknowledgments}

\end{document}